\documentclass[12pt]{article}
\usepackage{makeidx}
\usepackage{amssymb}
\usepackage{amsfonts}
\usepackage{amsmath}
\usepackage[T1]{fontenc}
\usepackage{graphicx}
\usepackage{setspace}
\usepackage{authblk}
\usepackage[left=2cm,top=2cm,right=2cm]{geometry}
\usepackage{units}
\usepackage{color}

\begin{document}
\title{
\begin{flushleft}
\normalsize{Updated talk, given by A.E.Shabad at the XVI Lomonosov Conference
held at Moscow State University, August 2013,
to appear in Proceedings, World Scientific, Singapore, edited by A.I. Studenikin}
\end{flushleft}
\vspace{10pt}
\textbf{Nonlinearity in Electro- and Magneto-statics with and without External Field}}
\author[1]{T. C. Adorno\thanks{tadorno@usp.br}}
\author[1]{Caio Costa\thanks{caiocostalopes@usp.br}}
\author[1]{D. M. Gitman\thanks{gitman@dfn.if.usp.br}}
\author[2]{A. E. Shabad\thanks{shabad@lpi.ru}}
\affil[1]{\textit{Instituto de F\'{\i}sica, Universidade de S\~{a}o Paulo, Caixa Postal 66318, CEP 05508-090, S\~{a}o Paulo, S.P., Brazil;}}
\affil[2]{\textit{P. N. Lebedev Physics Institute, Moscow, Russia}}

\maketitle

\onehalfspacing

\begin{abstract} Due to the nonlinearity of QED, a static
charge becomes a magnetic dipole if placed in a magnetic field.
Already without external field, the cubic Maxwell equation for the
field of a point charge has a soliton solution with a finite field
energy. Equations are given for self-coupling dipole moments. Any
theoretically found value for a multipole moment of a baryon or a
meson should be subjected to nonlinear renormalization.
\end{abstract}

\section{Introduction}
In this talk we give an overview of the results presented in
Refs.\cite{GitSha2012}--\cite{CosGitSha2013a}.

The nonlinear Maxwell equations truncated at the third power of
their expansion in the expectation value of the electromagnetic
field $a^{\tau}\left(  x\right)$ over the blank vacuum or the vacuum
with a background field have the form
\begin{equation}
 j_{\mu}\left(  x\right)+j_{\mu}^{\rm nl}\left(  x\right)  =\left[
\Box\eta_{\mu\tau}-\partial_{\mu }\partial_{\tau}\right]
a^{\tau}\left(  x\right)+\int{d}^{4}y\Pi_{\mu\tau}\left( x,y\right)
a^{\tau}\left(  y\right)\label{Maxwell} \end{equation} where
$\Box=\partial_{0}^{2}-\nabla^{2}$. The four-vector $j_{\mu}\left(
x\right)$ is the current to the field $a^{\tau}\left(  x\right),$
whereas $j_{\mu}^{\rm nl}\left( x\right)$ is the nonlinearly induced
current, depending in its turn on the field:
\begin{equation}
 j_{\mu}^{\rm nl}\left( x\right)=
 -\frac{1}{2}\int{d}^{4}y{d}^{4}u\Pi_{\mu\tau\sigma }\left(  x,y,u\right)  a^{\tau}\left(  y\right)
a^{\sigma}\left(  u\right)
\nonumber\\
  -\frac{1}{6}\int{d}^{4}y{d}^{4}u{d}^{4}v%
\Pi_{\mu\tau\sigma\rho}\left(  x,y,u,v\right)  a^{\tau}\left(
y\right) a^{\sigma}\left(  u\right)  a^{\rho}\left(
v\right)\label{nonlinear current1}
\end{equation}
The second-, third- and fourth-rank polarization tensors
$\Pi_{\mu\tau},$ $\Pi_{\mu\tau\sigma }$ and
$\Pi_{\mu\tau\sigma\rho}$ are responsible for the linear, quadratic
and cubic responses of the vacuum to the applied electromagnetic
field $a^{\tau}$. These tensors are defined as the second, third and
fourth variational derivatives of the effective action with respect
to the fields. They depend on the background field if the latter is
kept nonzero after these derivatives are calculated. When the field
strength of the background is time- and space-independent, the
polarization tensors of all ranks depend on the differences of the
coordinates.

Nontrivial results concerning solutions of the nonlinear Maxwell
equations (\ref{Maxwell}, \ref{nonlinear current1}) may be already
obtained provided we confine ourselves to the simplest
approximation, which stems from the effective action  $ \Gamma =\int
{{L}}\left( x\right) d^{4}x$ taken as a local functional, of the scalar $\left( {F}%
\right) $ and pseudoscalar $\left( {G}\right) $ field invariants,
for instance exemplify it as the Heisenberg-Euler action in QED, or
the Born-Infeld action beyond it. The local approximation is
applicable to the fields, slowly varying in time and space. When QED
is concerned, the corresponding space-time scale is determined by
the electron Compton length $m^{-1}$. The truncation of the exact
Maxwell equations at the third power of the field done above
restricts their applicability to not too strong fields, but still
stronger than the ones usually treated within the linear
approximation $j_{\mu}^{\rm nl}=0$. In QED the measure of strength
of the field is the value $m^2/e$. The whole approach is aimed to
cover strong electromagnetic fields in the vicinity of elementary
particles, although we cannot yet come too close to them owing to
the above restrictions.

 We shall consider here static fields of charges and of electric and
 magnetic dipoles. In Section 2 we shall deal with external constant magnetic
 field background and omit the cubic term in (\ref{nonlinear current1}).
 In Section 3 there will be no background, hence $\Pi_{\mu\tau\sigma
 }=0$ owing to the Furry theorem. So we shall face a cubic equation.
 \section{Magnetic moment of a spherical charge in a static and homogeneous magnetic background
 \cite{GitSha2012}, \cite{AdoGitSha2013}, \cite{AdoGitSha2014}}
Here we disregard $\Pi_{\mu\tau\sigma\rho}$. The third-rank tensor
is expressed in terms of the second and third derivatives
${L}_{FF}$, ${L}_{{GG}}$, ${L}_{{FGG}}$ of the effective Lagrangian
$L$ with respect to the two electromagnetic
field invariants taken at ${G=}$ $0,$ $2{F=}$$B^{2}$ = \emph{%
const.} They depend only on the background time- and space
independent magnetic field $B=|\mathbf{B}|$.
Then the current (\ref{nonlinear current1})in the totally static field-configuration is \cite{GitSha2012}:%
\begin{equation}
j^{\mathrm{nl}}_{0}\left( \mathbf{x}\right) =0\,,\ \ {j_i}^{\mathrm{nl}%
}\left( \mathbf{x}\right) =\epsilon_{ikj} \nabla_k\eta_j
\left( \mathbf{x}\right),  \label{nonlinear current2}
\end{equation}
\begin{equation}
\eta_{i}\left( \mathbf{x}\right) =\frac {B_i}2E^2{{L}}_{{FF}}%
-\frac{B_i}2\left( {\mathbf{B}}\cdot {\mathbf{E}}\right)^2{L}_{{FGG}}%
-{{E}_i}\left( {\mathbf{B}\cdot\mathbf{E}}\right){L}_{{GG}}%
\,. \label{auxiliary field}
\end{equation}
Here $\mathbf{E}$ is the static applied field contained in the
vector-potential in the right-hand side of (\ref{nonlinear
current1}).

 We treat here the nonlinearity as perturbation. Then $E$ is
 understood as the electric field created by the static charge $j_0\neq0$, $j_i=0$  via
 the equation (\ref{Maxwell}), wherein the nonlinear current (\ref{nonlinear
 current1}) is set equal to zero, $j^{\rm nl}_\mu=0.$ Then the
 magnetic field produced by this electric field results from (\ref{nonlinear
 current2}) and (\ref{auxiliary field}) via the Maxwell equations
 (\ref{Maxwell}) without the second term in the r.-h. side (it would
 give a contribution of higher order; see, however,
 \cite{AdoGitSha2014} for its full account) to be:\begin{equation}
h_{i}\left( \mathbf{x}\right)=\left(  \delta_{ik}%
-\frac{\nabla_{i}\nabla_{k}}{\nabla^{2}}\right) \eta_{k}\left(
\mathbf{x}\right) =\eta_{i}\left( \mathbf{x}\right) +%
\frac{\partial _{i}\partial _{k}}{4\pi }
\int d^{3}y\frac{%
\eta_{k}\left( \mathbf{y}\right) }{\left\vert \mathbf{x}-\mathbf{y}%
\right\vert }\,.
  \label{magfield}
\end{equation}%

Consider the magnetic field, which is the response of the vacuum to
the applied electric field, whose vector potential is chosen
to be the following smooth central-symmetrical Coulomb-like function%
\begin{equation}\label{potential}
a_{0}\left( r\right) =\left(-\frac{Ze}{8\pi R^{3}}r^{2}+\frac{3Ze}{%
8\pi R}\right) \theta \left( R-r\right) +\frac{Ze}{4\pi r}\theta
\left( r-R\right) \,,\,\,r=\left\vert \mathbf{x}\right\vert.
\end{equation}%
Here $\theta \left( z\right) $ is the step function,
defined as
\begin{equation}
\theta \left( z\right) =\left\{
\begin{array}{c}
1\,,\ \ z>0\,, \\
0\,,\ \ z<0\,%
\end{array}%
\right. \,.
\end{equation}%
Disregarding the higher-order effect of the linear electrization we
state that (\ref{potential}) is the potential of the charge
distributed inside the sphere with the radius $R$ with the constant
density $\rho \left( r\right) =\left( \frac{3}{4\pi
}\frac{Ze}{R^{3}}\right) \theta \left( R-r\right)$. The long-range
contribution of (\ref{magfield}) calculated with the electric field
contained in (\ref{potential}) (see \cite{AdoGitSha2013} and
\cite{AdoGitSha2014} for its explicit form),
$h_{i}^{\mathrm{LR}}\left( \mathbf{x}\right) $, behaves like a
magnetic field generated by a magnetic
dipole:%
\begin{equation}
h_{i}^{\mathrm{LR}}\left( \mathbf{x}\right) =\frac{3\left(
\mathbf{x}\cdot \mathbf{M }\right) x_{i}}{r^{5}}-\frac{M _i}{r^3}\,,
\label{18.9}
\end{equation}%
with $\mathbf{M}$ being the equivalent magnetic dipole moment, given
by%
\begin{equation}
M _{i}=\left(\frac{Ze}{4\pi}\right)^2\frac{1}{5R}\left( 3{L}_{%
{FF}}-2{L}_{{GG}}-B^{2}{L}_{{FGG}%
}\right) B_{i}\,.  \label{18.10}
\end{equation}
The extension beyond the spherical symmetry may be found in
\cite{AdoGitSha2014}.
\section{Cubic self-interaction of electro- and magneto-static fields in blank vacuum
\cite{CosGitSha2013}, \cite{CosGitSha2013a}}
In this section no
background field is present, hence $\Pi_{\mu\tau\sigma }=0$ in
(\ref{nonlinear current1}). Besides, the principle of correspondence
with the classical Faraday-Maxwell electromagnetism requires that
also $\Pi_{\mu\tau}=0$ within the local limit dealt with here,
because in this limit the quantum theory is normalized to the
classical one. Consequently, there is no linear polarization, and
inductions and field strengths are the same in the blank vacuum.

 Unlike the previous section, now the
nonlinearity in the Maxwell equation (\ref{Maxwell}) will not be
taken as small, but treated seriously. In the two subsections below
we include only the static cases, where, besides, the field $a_\mu$
in the nonlinear current (\ref{nonlinear current1}) carries either
only electric, $E,$ or only magnetic, $B,$ field, not the both
simultaneously. Then this current is calculated to
be\begin{equation}\label{blankvaccur}
 j_{0}^{\mathrm{nl}}\left(  x\right)  =\frac{1}{2}{L}%
_{{FF}}{\partial_i}\left[  \left( {B}%
^{2}-{E}^{2}\right) {E_i}\right],\qquad
  {j_i}^{\mathrm{nl}}\left(  x\right)  =-\frac{1}{2}{L}%
_{{FF}}\partial_j \left[  \left(  {B}^{2}-{E}^{2}\right) {B_k}\right]\epsilon_{ijk}%
. \end{equation} In the present section the derivative
${L}_{{FF}}\equiv\gamma$ is understood as taken at $F=G=0$.

\subsection {Self-coupling of a charge. Finiteness
of the point-charge electrostatic field-energy}
 Let there be a point charge $e$ placed at the origin $r=0$.
We are looking for a spherically symmetric solution of the Maxwell
equation (\ref{Maxwell}) with $B=0$, which, given the nonlinear
current (\ref{blankvaccur}), takes the form (we denote $\gamma\equiv
L_{FF}$ for brevity)\begin{equation} {\nabla}\left[ \left(
1+\frac{\gamma}{2}E^{2}\right) \mathbf{E}\right] =0,
\label{Electrostatic equation1}
\end{equation} valid everywhere outside
the origin $\mathbf{x}=0$, since $j_0=0$ there.
At large $%
r $ the standard Coulomb field of the point charge $e$%
\begin{equation}\label{Coulomb}
\frac{e}{4\pi r^{2}}\frac{\mathbf{x}}{r},
\end{equation}  should be implied as the boundary condition.
Then with the spherically symmetric Ansatz $
 E(r)\frac{\mathbf{x}}{r}=\mathbf{E}\left(
\mathbf{x}\right)$  equation (\ref{Electrostatic equation1}) is
solved as
\begin{equation}
\left( 1+\frac{\gamma}{2}E^{2}(r)\right) E\left( r\right)
=\frac{e}{4\pi r^{2}}.  \label{Cubic equation}
\end{equation} This cubic equation is readily solved by the Cardan
formula (see \cite{CosGitSha2013a} for the explicit representation),
but the most important thing about its solution is clear without
solving it: at short distances $r\to\infty$ the field $E$ also
infinitely grows, hence one can neglect the unity in (\ref{Cubic
equation}) to immediately obtain \begin{equation}
E\left( r\right) \sim \left( \frac{e}{2\pi{\gamma}}%
\right) ^{\frac{1}{3}}\left( \frac{1}{r}\right) ^{\frac{2}{3}},
\label{Enear origin}\end{equation}The behavior (\ref{Enear origin})
of the electrostatic field , produced by the point charge $e$ via
the nonlinear field equations (\ref{Maxwell}), is essentially less
singular in the vicinity of the charge than the standard Coulomb
field $\frac{e}{4\pi r^{2}}.$

Let us see that this suppression of the singularity is enough to
provide convergence of the integrals giving the energy of the field
configuration that solves equation (\ref{Electrostatic equation1}).
To this end note that on the subclass of electromagnetic field we
are considering here, the equations of motion (\ref{Maxwell}), or,
equivalently, (\ref{Electrostatic equation1}) are generated by the
quartic Lagrangian\begin{equation}-{F}\left( x\right)+L=
-{F}\left( x\right) +\frac{\gamma }{2}\left({F(}%
x)\right) ^{2}.  \label{L}
\end{equation}%
With this Lagrangian, the energy density calculated on
spherically-symmetric electric field configuration following the
Noether theorem is\begin{equation}\label{theta00}
\Theta^{00}=\frac{E^{2}}{2}+\frac{3\gamma E^{4}}{8}.
\end{equation}The behaviour (\ref{Enear origin}) provides the
ultraviolet, near $|\mathbf{x}|=0$, convergence of the electrostatic
field energy $\int\Theta^{00}d^3x$ of the point charge. As for the
convergence of this integral at $|\mathbf{x}|\to\infty,$ it is
provided by the standard long-range Coulomb behaviour
(\ref{Coulomb}) of the solution  to equation (\ref{Electrostatic
equation1}) obtained by neglecting the second term inside the
bracket as compared to the unity in the far-off region.

The explicit use of the Cardan formula in (\ref{theta00}) allows to
calculate the integral for the field energy. If the value
\begin{equation}
L_{FF}=\frac{e^{4}}{45\pi^{2}m^{4}},  \label{gamma}
\end{equation} where $e$ and $m$ are the electron charge and mass,
 is calculated referring to the Euler-Heisenberg
effective Lagrangian of QED and substituted for $\gamma,$ the result
for the "rest mass of the electron," understood as a point charge,
is about twice the true electron mass:\begin{equation}
\int\Theta^{00}{d}^{3}x=2.09m.  \label{mass}
\end{equation}

The conclusion about finiteness of the electrostatic field energy of
a point charge is readily extended \cite{CosGitSha2013a} to the
nonlinear electrodynamics with the effective Lagrangian being
any-power polynomial of the field invariants, thereby also to QED
truncated at any finite term of its Taylor expansion in powers of
the field in place of (\ref{Maxwell}).
\subsection {Self-coupling of magnetic and electric dipoles}
Consider first a magnetic dipole. This means that only $B$ is kept
in the nonlinear current (\ref{blankvaccur}), hence $j_0^{\mathrm
nl}=0.$ As for the nonlinear 3-current, it is expressed as
\begin{equation}
{j_i}^{\mathrm{nl}}\left( \mathbf{ x}\right)  =\epsilon_{ijk} {\nabla_j}%
\eta_k\left(  \mathbf{x}\right),\qquad\eta_i\left(  \mathbf{x}\right)  =-\frac{1}{2}%
{L}_{{FF}}{B_i}\left(  \mathbf{x}\right) B^{2}\left(
\mathbf{x}\right)  \label{Cubic magnetic field}%
\end{equation}in terms of the auxiliary magnetic field $\eta$
analogous to (\ref{auxiliary field}). Eq. (\ref{magfield}) is again
valid for the magnetic field induced by the nonlinear current, this
time without the reservations made in the previous section about the
disregard of the linear magnetization. This field is to be added to
the initial magnetic field $\mathbf{h}^{\mathrm{nl}}$ (linearly
produced by the current $\mathbf{j}$) to make the total resulting
magnetic field $\mathbf{h}^{\mathrm{tot}}=\mathbf{h}^{\mathrm
{nl}}+\mathbf{h}.$

 Let there be a sphere with the
radius $R,$ and a time-independent current
$\mathbf{j}(\mathbf{x})$ concentrated on its surface:%
\begin{equation}\label{dipcur}
{\mathbf{j}(\mathbf{x}})%
=\frac{\mathbf{{M}}^{(0)}\times {\mathbf{x}}}{r^{4}}\delta(r-R).%
\end{equation}
Here $\mathbf{{M}}^{(0)}$ is a constant vector directed, say, along
the axis 3. The current density ($\ref{dipcur}$) obeys the
continuity condition $\nabla\mathbf{j}$ $=0$, its flow lines are
circular in the planes parallel to the plane (1,2). The magnetic
field produced by this current via the Maxwell equation
$\nabla\times\mathbf{h}^{\mathrm{lin}}(\mathbf{x)}=$
$\mathbf{j}(\mathbf{x})$ is\begin{equation}
{\mathbf{h}^{\mathrm{lin}}(\mathbf{x}})=\theta\left( R-r\right)
\frac{2\mathbf{{M}}^{(0)}}{R^{3}} +\theta\left( r-R\right)\left(
-\frac{\mathbf{{M}}^{(0)}}{r^{3}%
}+3\frac{\mathbf{x}\cdot\mathbf{{M}}^{(0)}}{r^{5}%
}\mathbf{x}\right)   . \label{Blin}%
\end{equation} Outside the sphere this is the
magnetic dipole field
with the constant vector density $\mathbf{{M}^{(0)}}$  playing the role of its magnetic moment. Using this expressioin in the r.-h. side of  Eq. (\ref{magfield}), after a lengthy calculation the nonlinear correction $h$ to the field (\ref{Blin} )of the magnetic dipole(\ref{dipcur}) was obtained in \cite{CosGitSha2013} both inside and outside the sphere. At large distances the resulting field reproduces the original magnetic dipole behaviour:
\begin{equation}
\left.  \mathbf{h}^{\mathrm tot}\left(  \mathbf{r}\right)  \right\vert _{r>>R}%
={\mathbf{h}}^{\mathrm{lin}}\left(  \mathbf{r}\right)  \left(  1-\frac{7}%
{5}{L}_{{FF}}\frac{{M}^{(0)2}}{R^{6}}\right)  .
\label{Final correction}%
\end{equation}
Once we want to treat the nonlinearity seriously, and not just as a perturbation, we should for self-consistency demand that the magnetic field forming the nonlinear current (\ref{Cubic magnetic field}) be not (\ref{Blin}), but its final result, which is again the magnetic dipole field, but with the bare magnetic moment $\mathbf{{M}^{(0)}}$ replaced by the final magnetic moment to be denoted as $\mathbf{{M}}$. Then in the long range for the total field we obtain
\begin{equation}
-\frac{\mathbf{M}}{r^{3}}+3\frac{\mathbf{x}\cdot\mathbf{M}}{r^{5}%
}{\mathbf{x}}=-\frac{\mathbf{{M}}^{(0)}}{r^{3}}+3\frac{\mathbf{x}\cdot\mathbf{{M}}^{(0)}}{r^{5}%
}{\mathbf{x}}-\left(\frac{\mathbf{{M}}}{r^{3}}+3\frac{\mathbf{x}\cdot\mathbf{{M}}}{r^{5}%
}\mathbf{x}\right)
  \left(\frac{7}%
{5}{L}_{{FF}}\frac{{M}^{2}}{R^{6}}\right) .
\end{equation}   From this the equation for self-coupling of the magnetic moment follows to be:
\begin{equation}{\mathbf{M}}\left(1+ \frac{7}%
{5}{L}_{{FF}}\frac{{M}^{2}}{R^{6}}\right)=\mathbf{M}^{(0)}\end{equation}
Analogous equation for the electric moment is\begin{equation} {\mathbf{p}}\left(1+\frac 1{10}L_{FF}\frac{p^2}{R^6}\right)=\mathbf{p}^{(0)}.\end{equation}
\section*{Acknowledgments}
Supported by CAPES and FAPESP under grants 2013/00840-9 and 2013/16592-4, by CNPq,  by RFBR under the Project
11-02-00685-a and by the project 2.3684.2011 of TSU.

\end{document}